\documentclass[prl,twocolumn,amsmath,amssymb]{revtex4}

\usepackage{graphicx}
\usepackage{dcolumn}
\usepackage{bm}

\begin{document}


\title{On the density matrix of nonequilibrium steady-state statistical mechanics}

\author{Takafumi Kita}
\affiliation{Division of Physics, Hokkaido University, Sapporo 060-0810, Japan}

\date{\today}

\begin{abstract}
This paper derives a density matrix of the steady-state statistical mechanics
compatible with the steady-state thermodynamics proposed 
by Oono and Paniconi
[Prog.\ Theor.\ Phys.\ Suppl.\ {\bf 130},\ 29 (1998)].
To this end, we adopt three plausible basic assumptions for uniform 
steady states: (i) equivalence between any two subsystems of the total,
(ii) statistical independence between any two subsystems,
and (iii) additivity of energy.
With a suitable definition of energy,
it is then shown that uniform steady states driven by mechanical forces 
may be described by the Gibbs distribution.
\end{abstract}


\maketitle

Constructing thermodynamics and statistical mechanics 
far from equilibrium is undoubtedly a major
goal yet to be achieved in modern condensed matter physics.
One strategy toward this may be to seek a way to extend
the well-established equilibrium framework to nonequilibrium systems.
Numerous efforts have been made along this 
line\cite{dGM62,Fitts62,Prigogine71,Zubarev74,Landauer78,Keizer87,Jou93,Eyink96}.
However, most of them starts from the local-equilibrium hypothesis
which may not be justified for systems far from equilibrium.
Recently, Oono and Paniconi\cite{OP98} presented a new approach
restricting their attention to 
nonequilibrium time-independent states.
A key ingredient lies in the removal of
``house keeping heat rate'' $Q_{\textrm{hk}}$ which is generated in the system 
as dissipation
to be carried away eventually by some microscopic degrees of freedom.
Using the ``excess heat rate'' $Q_{\textrm{ex}}$ defined by subtracting
$Q_{\textrm{hk}}$
and connecting points of steady ``state space'' in a well-defined way, 
they have constructed a thermodynamic framework
named ``steady-state thermodynamics'' (SST)
which is quite analogous to the equilibrium one.
Indeed, the energy and the Helmholtz free energy
are defined there in the same way as equilibrium thermodynamics
by merely introducing extra extensive variables characteristic of
the relevant steady state.

Naturally, an amount of their effort was directed towards an unambiguous
definition of entropy which serves as the potential for nonequilibrium systems
to determine their evolution and stability criteria.
It should be noted that they are completely different from
the variational principle on entropy production rate established
by Onsager near equilibrium\cite{Onsager31} and taken up by Glansdorff
and Prigogine\cite{Prigogine71}
as a candidate for the general criteria 
in nonequilibrium systems.
Indeed, entropy and entropy production rate are different in dimension
so that the two variational principles cannot be compatible with 
each other.
It is intuitively plausible to expect that,
once the steady dissipation is thrown away into 
``hidden degrees of freedom''\cite{Callen},
we may have well-defined energy and entropy which play the
same role as the equilibrium thermal physics.

Efforts have been made thereafter to test SST
based on microscopic stochastic models\cite{Sekimoto98,Hatano99,HS01,Shibata00}.
However, those stochastic models are so designed as to settle down in
equilibrium, it is not clear whether the results from them 
are really relevant to far-from-equilibrium states,
as recently pointed out by Sasa and Tasaki\cite{ST01}.

We here take an alternative approach to seek for a statistical 
density matrix which is compatible with SST
of Oono and Paniconi\cite{OP98}.
There are a couple of merits in this approach.
First, once such a density matrix is obtained, we may perform 
microscopic calculations on nonequilibrium
steady states which could be tested by experiments.
Second, this statistical mechanics may make clearer the concepts 
of SST such as ``state space,''
``energy,'' and ``entropy.''
Hopefully the attempts will also be helpful to clarify the conditions under which 
SST holds.

We restrict ourselves to the cases of mechanical external
perturbations where an unambiguous treatment is possible.
As is well known, the thermal perturbations are difficult to handle even
within the linear-response regime\cite{KTH}, so that we leave those cases
for a future consideration.

We specifically consider electrons in a metal with impurities
which is embedded in a uniform electric field ${\bf E}$;
this electric field sustain uniform current ${\bf J}$
in the system.
First, let us review a conventional way to describe the system:
It is convenient to express the electric field
in terms of the time-dependent vector potential
${\bf A}(t)$ as ${\bf E}=
-\frac{\partial {\bf A}(t)}{\partial t}$.
The corresponding Hamiltonian is given in units of $\hbar=1$
by
\begin{eqnarray}
{\cal H}(t)=\!\int \!\psi^{\dagger}({\bf r})\left\{\!
\frac{[-i\bm{\nabla}\!-\! e{\bf A}(t)]^{2}}{2m}\!+\! V_{\rm imp}({\bf r})
\!-\!\mu\right\}\! \psi({\bf r}) \, d{\bf r} 
\nonumber \\
\label{H_0}
\end{eqnarray}
with $e\hspace{0.5mm}(<\! 0)$ the electron charge, $m$ the electron mass, 
$V_{\rm imp}$ the impurity potential,
and $\mu$ the chemical potential.
The spin degrees of freedom and the electron-electron interaction 
are suppressed for simplicity.
A general advantage of using time-dependent Hamiltonians for nonequilibrium states
is that we can definitely identify the input power as
$\langle\frac{\partial {\cal H}(t)}{\partial t}\rangle$.
Here, electrons are accelerated along ${\bf E}$ to acquire the input
power, but then scattered by $V_{\rm imp}$ resulting in their momentum
relaxation. Finally, the energy relaxation occurs, i.e.,
the electrons interact with some other microscopic degrees 
of freedom such as phonons, and the extra energy
accumulated in the electrons are eventually carried away from the system.
However, this last stage cannot be described by the density matrix 
$\rho(t)$
obtained by solving the quantum Liouville equation\cite{Zubarev74,KTH}:
\begin{eqnarray}
i\frac{\partial}{\partial t}\rho(t)=[{\cal H}(t),\rho(t)]\, .
\label{QL}
\end{eqnarray}
Indeed, its formal solution is given by
\begin{eqnarray}
\rho(t)=&&\hspace{-3mm}{\cal S}(t)\rho_{0}{\cal S}^{\dagger}(t)\, ,
\label{rho}
\end{eqnarray}
where $\rho_{0}$ is the equilibrium density matrix at $t=t_{0}$
when ${\bf E}$ is switched on,
and ${\cal S}(t)$ is defined by
\begin{eqnarray}
{\cal S}(t)=T \exp\!\left[-i\int_{t_{0}}^{t}{\cal H}(t')dt'\right] \, .
\label{S-matrix}
\end{eqnarray}
Equation (\ref{rho}) forms a basic starting point 
for the linear-response theory
and the fluctuation-dissipation theorem\cite{Zubarev74,KTH}.
Even beyond the linear-response regime, 
it provides a general framework 
to perform nonlinear nonequilibrium calculations\cite{Keldysh64,Rammer86}.
However, since $\rho(t)$ develops purely mechanically from
equilibrium $\rho_{0}$ with no thermal contact,
energy relaxation processes are absent in Eq.\ (\ref{QL}).
Thus, exact calculations based on Eq.\ (\ref{QL}) predict 
that the energy of the electron system grows towards infinity
as time goes by.
It might be possible to add terms responsible
for the energy relaxation to Eq.\ (\ref{rho}) \cite{Zubarev74},
but there seems to be an ambiguity
in this procedure especially in the nonlinear region.
Thus, the quantum Liouville equation may not be
a good starting point to obtain the correct 
steady-state density matrix
far from equilibrium.
Even more difficult will be to perform calculations 
of including microscopic degrees of freedom 
which transport the extra electron energy out of the system,
because whether the combined system settles down into a steady state by itself
is not entirely clear.

To seek for the density matrix, we here proceed differently
based on an argument which assumes the existence of a uniform steady state
from the beginning and 
which fully relies upon the uniformity of the system. 
In the end, the equilibrium density matrix can be derived 
from: (i) equivalence between any two subsystems of the total,
(ii) statistical independence between any two subsystems,
and (iii) additivity of energy.
These are essential ingredients of the concept of entropy.
Since (i) and (ii) are
also characteristic of any uniform steady states,
we may expect to have a well-defined entropy
by a suitable definition of energy for each subsystem.
An extension to nonuniform steady states may be performed through
the usual procedure of dividing the total system 
into small uniform cells\cite{dGM62,Fitts62}.

\begin{figure}
\includegraphics[width=0.9\linewidth]{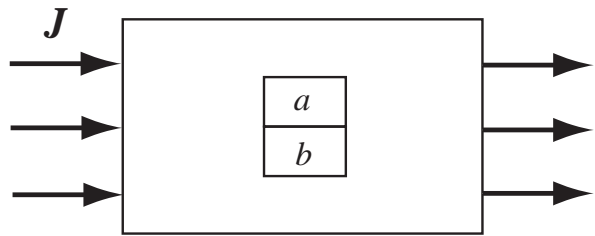}%
\caption{The total system with
a uniform current and its subsystems.}
\end{figure}

Consider a subsystem $a$ of the total where steady current
${\bf J}$ is present (see Fig.\ 1).
In this subsystem, there is equal amount of energy input and dissipation
per unit time,
the latter being thrown out of the subsystem
by some microscopic degrees of freedom
not considered explicitly.
Following the philosophy of SST, 
we disregard this energy flow,
include the carrier of the dissipations in 
``hidden degrees of freedom''\cite{Callen},
and consider the fluctuations caused by the energy flow and
those through the wall of the subsystems together.
Now, the system is characterized by the uniform static current ${\bf J}$,
whose ``energy'' may be calculated by the static ``Hamiltonian:''
\begin{eqnarray}
{\cal H}^{(a)}=&&\hspace{-3mm}\int_{V_{a}} \!\psi^{\dagger}({\bf r})\!\left[
\frac{(-i\bm{\nabla} -{\bf p}_{0})^{2}}{2m}+V_{\rm imp}({\bf r})-\mu\right] 
\!\psi({\bf r}) \, d{\bf r} \, ,
\nonumber \\
\label{H2}
\end{eqnarray}
where ${\bf p}_{0}$ is a variable conjugate to ${\bf J}$.
From now on we can follow the argument of equilibrium statistical
mechanics to derive the density matrix $\rho^{(a)}$ of the subsystem $a$:
Since it is time-independent by assumption,
$\rho^{(a)}$ commutes with ${\cal H}^{(a)}$, so that it
is a function of the eigenvalue ${\cal E}_{\nu}^{(a)}$ of ${\cal H}^{(a)}$.
From (ii) the statistical independence, $\rho^{(a+b)}=\rho^{(a)}\rho^{(b)}$,
and (iii) additivity of energy, ${\cal E}^{(a+b)}={\cal E}^{(a)}+{\cal E}^{(b)}$,
it follows that\cite{LL}
\begin{eqnarray}\ln\rho_{\nu}^{(a)}=\alpha^{(a)}
-\beta{\cal E}_{\nu}^{(a)} \, ,
\label{lnrho}
\end{eqnarray}
where $\alpha^{(a)}$ and $\beta$ are some constants.
Normalizing $\rho^{(a)}$, we have
\begin{eqnarray}
\rho^{(a)}_{\nu}=\frac{{\rm e}^{-\beta {\cal E}_{\nu}^{(a)}}
}{\displaystyle\sum_{\mu}{\rm e}^{-\beta {\cal E}_{\mu}^{(a)}}} \, ,
\label{rhoa}
\end{eqnarray}
where higher probability for lower ${\cal E}$ means $\beta>0$.
Thus, we have reached a candidate for the steady state density matrix 
in the presence of a uniform current $\bf J$.

We now see that Eq.\ (\ref{rhoa}) corresponds to the state 
of maximum entropy,
following exactly the argument of equilibrium case\cite{LL}:
Entropy is defined by $S^{(a)}\equiv -{\rm Tr}\rho^{(a)}\ln \rho^{(a)}
\equiv -\left<\ln\rho^{(a)}\right>$ 
and calculated by using Eqs.\ (\ref{H2}) and (\ref{lnrho}) as
\begin{eqnarray}
S^{(a)} &&\hspace{-4mm}= - \alpha^{(a)} +\beta\langle{\cal E}^{(a)}\rangle 
\nonumber \\
&&\hspace{-4mm}=-\alpha^{(a)}+\beta(\langle E^{(a)}\rangle-
\mu' \langle N^{(a)}\rangle
-{\bf p}_{0} \cdot\langle {\bf J}^{(a)}\rangle)\, ,
\label{Sa}
\end{eqnarray}
where $E^{(a)}$ is the energy of the subsystem, 
$\langle {\bf J}^{(a)}\rangle\equiv -
\langle\partial {\cal H}^{(a)}/\partial {\bf p}_{0}\rangle$, and 
$\mu'\equiv \mu+\frac{p_{0}^{2}}{2m}$.
We also notice that $ -\langle\ln\rho^{(a)}\rangle= 
-\ln\rho^{(a)}(\langle{\cal E}\rangle)$.
On the other hand, Eq.\ (\ref{rhoa}) tells us that the probability
is the same for all degenerate states and equal
in the thermodynamic limit 
to the inverse of the number of states
around $\langle{\cal E}^{(a)}\rangle$.
Hence Eq.\ (\ref{Sa}) corresponds to the entropy maximum 
for the energy $\langle E^{(a)}\rangle$, the electron number 
$\langle N^{(a)}\rangle$, 
and the current $\langle {\bf J}^{(a)}\rangle$.
Finally, we may perform the Legendre transformation 
$G^{(a)}\equiv\langle 
{\cal H}^{(a)}\rangle+{\bf p}_{0}\cdot \langle {\bf J}^{(a)}\rangle$
to obtain the desired free energy $G^{(a)}$ as a function of $\beta$, $\mu'$,
and $\langle {\bf J}^{(a)}\rangle$.

The above consideration can be applied to any uniform steady states
driven by mechanical forces.
Indeed, we only have to introduce in the 
Hamiltonian a variable [such as ${\bf p}_{0}$ in Eq.\ (\ref{H2})]
which is conjugate to the
expectation value (such as ${\bf J}$ in the above consideration)
characterizing the steady state.

Several comments are in order. 
First, the consideration here puts aside completely
the driving force ${\bf E}$ and the corresponding dissipations.
Thus, this density matrix cannot say anything about 
the ${\bf E}$-${\bf J}$ relation, but only identifies the steady state
for a given ${\bf J}$.
However, this ${\bf E}$-${\bf J}$ relation may be obtained by connecting
the results through the linear-response calculations for each ${\bf J}$.
Second, the above argument starts from three plausible 
basic assumptions on uniform steady states:
(i) equivalence between any two subsystems of the total,
(ii) statistical independence between any two subsystems,
and (iii) additivity of energy.
Hence we may expect that the derived Gibbs distribution (\ref{rhoa}) 
is the correct distribution for those steady states.
Indeed, there are several stochastic models which realize
the Gibbs distribution in nonequilibrium steady states.\cite{OY87,Takesue87,
Yeung89}
However, it should be noted that
the quantity $\beta$ here includes the fluctuation of energy
due to dissipations as well as that through the walls among subsystems.
If the former contribution is negligible, we may expect that $\beta$ 
is the same as the equilibrium value.
Experimentally, there remains a basic problem on how to
measure the value of $\beta$ for the steady states.
Third, the above consideration have surely made the concept of ``state space''
in SST clearer.
The postulated second-law of SST may be proved now by using 
Eq.\ (\ref{rhoa}) and the Jarzynski identity\cite{Jarzynski}.
Fourth, the subsystem considered here may
contain structures small compared with the subsystem size
such as the vortex lattice structures.

There are many interesting nonequilibrium phenomena
found in semiconductors.\cite{Scholl}
Also, Stoll {\em et al}.\ found steps in the $I$-$V$ characteristics in
the vortex state of superconducting Nd$_{2-x}$Ce$_{x}$CuO$_{y}$;
this first-order-like transition may be described by
the nonequilibrium free energy derived above,
as increasing ${\bf E}$ causes the density-of-states (DOS) change 
in the electron system.

I am grateful for Koji Nemoto for an enlightening discussion.





\begin{references}

\bibitem{dGM62} S. R. de Groot and P. Mazur,
{\it Non-Equilibrium Thermodynamics} (North-Holland, Amsterdam, 1962).

\bibitem{Fitts62} D. D. Fitts,
{\it Nonequilibrium Thermodynamics: a Phenomenological Theory of 
Irreversible Processes in Fluid Systems} (McGraw-Hill, New York, 1962).

\bibitem{Prigogine71} P. Glansdorff and I. Prigogine,
{\it Thermodynamic Theory of Structure, Stability and Fluctuations} 
(Wiley-Interscience, London, 1971).

\bibitem{Zubarev74} D. N. Zubarev, {\it Nonequilibrium Statistical Thermodynamics}
(Consultants Bureau, New York, 1974).

\bibitem{Landauer78}R. Landauer, Phys. Rev. A{\bf 18}, 255 (1978);
Physica A{\bf 194}, 551 (1993).

\bibitem{Keizer87} J. Keizer, 
{\it Statistical Thermodynamics of Nonequilibrium Processes}
(Springer-Verlag, Berlin, 1987).

\bibitem{Jou93} D. Jou, J. Casas-V\'azques, and G. Lebon, 
{\it Extended Irreversible Thermodynamics}
(Springer-Verlag, Berlin, 1993).


\bibitem{Eyink96} G. L. Eyink, J. L. Lebowitz, and H. Spohn, 
J. Stat. Phys. {\bf 83}, 385 (1996).


\bibitem{OP98} Y. Oono and M. Paniconi, Prog. Theor. Phys. Suppl. {\bf 130}, 29 (1998).

\bibitem{Onsager31} L. Onsager, Phys. Rev. {\bf 37}, 405 (1931); {\bf 38}, 2265 (1931).

\bibitem{Callen}H. B. Callen, {\em Thermodynamics} (John Wiley, New York, 1960).

\bibitem{Sekimoto98} K. Sekimoto, Prog. Theor. Phys. Suppl. {\bf 130}, 17 (1998).

\bibitem{Hatano99} T. Hatano, Phys. Rev. E {\bf 60}, R5017 (1999).

\bibitem{HS01} T. Hatano and S. Sasa, Phys. Rev. Lett. {\bf 86}, 3463 (2001).

\bibitem{Shibata00} T. Shibata, cond-mat/0012404.

\bibitem{ST01} S. Sasa and H. Tasaki, cond-mat/00108365.

\bibitem{KTH} R. Kubo, M. Toda, and N. Hashitsume, {\it Nonequilibrium
statistical mechanics} (Springer-Verlag, Berlin, 1991).

\bibitem{Keldysh64} L. V. Keldysh, Zh. Eksp. Teor. Fiz. {\bf 47}, 1515 (1964)
[Sov. Phys. JETP {\bf 20}, 1018 (1965).

\bibitem{Rammer86}
For a review, see for example,  J. Rammer and H. Smith, Rev. Mod. Phys. {\bf 58}, 323
(1986); H. Haug and A.-P. Jauho, {\em Quantum Kinetics in Transport and Optics 
of Semiconductors} (Springer, Berlin, 1998) Chap 4.


\bibitem{LL} L. D. Landau and E. M. Lifshitz, {\em Statistical Physics}
(Pergamon, Oxford, 1980) \S 7.

\bibitem{OY87}Y. Oono and C. Yeung, J. Stat. Phys. {\bf 48}, 593 (1987).

\bibitem{Takesue87}S. Takesue, Phys. Rev. Lett. {\bf 59}, 2499 (1987).

\bibitem{Yeung89}C. Yeung, J. Stat. Phys. {\bf 55}, 357 (1989).

\bibitem{Jarzynski} C. Jarzynski, Phys. Rev. Lett. {\bf 78}, 2690 (1997).

\bibitem{Scholl} E. Sch\"oll, {\em Nonequilibrium Phase Transitions in Semiconductors}
(Springer-Verlag, Berlin, 1987)

\bibitem{Stroll99} O. M. Stroll, R. P. H\"ubener, S. Kaiser, and M. Naito,
Phys. Rev. B{\bf 60}, 12424 (1999).





\end{references}
\end{document}